# African rice cultivation linked to rising methane

Zichong Chen[1*], Nicholas Balasus[1], Haipeng Lin[1], Hannah Nesser[2], and Daniel J. Jacob[1]

[1]School of Engineering and Applied Sciences, Harvard University, Cambridge, MA, USA

[2]Jet Propulsion Laboratory, California Institute of Technology, Pasadena, CA, USA

*Correspondence to: Zichong Chen (zchen1@g.harvard.edu)

**Africa has been identified as a major driver of the current rise in atmospheric methane, and this has been attributed to emissions from wetlands and livestock. Here we show that rapidly increasing rice cultivation is another important source, and estimate that it accounts for 7% of the current global rise in methane emissions. Continued rice expansion to feed a rapidly growing population should be considered in climate change mitigation goals.**

Methane ($CH_4$) is a potent greenhouse gas responsible for 0.6°C global warming since the pre-industrial era[1]. Methane is emitted from many anthropogenic source sectors including the oil and gas supply chain, coal mining, livestock, waste management, and rice cultivation. The main natural source is wetlands. The accelerating rise in atmospheric methane since 2007 remains poorly understood and poses a major challenge for achieving the goal of the Global Methane Pledge (30% decrease in methane emissions by 2030; Ref[2]). Global methane emissions, estimated at 538-614 Tg $a^{-1}$, have been increasing at a rate of 2.8 ± 0.67 Tg $a^{-1}$ $a^{-1}$ over the 2006-2017 period[3,4]. Here we propose that rapidly increasing rice emissions in Sub-Saharan Africa (SSA) could be a significant but previously unrecognized contributor to the global methane trend.

Inverse top-down analyses of atmospheric methane observations have pointed to tropical Africa as a major driver of the current methane rise, attributing it to wetlands[5-8] or livestock[9]. Observations of the stable isotope $^{13}C$-$CH_4$ ($^{13}CH_4$) point to a dominant contribution from microbial sources in tropical Africa[10-11]. West Africa emerges from the inverse analyses as a major methane emission growth region[9,12,13]. A global inversion using Greenhouse Gases Observing Satellite (GOSAT) data for 2010-2018 inferred an increasing trend of 0.28 Tg $a^{-1}$ $a^{-1}$ in anthropogenic emissions from West Africa and found that livestock could only explain one third of that trend[9].

In response to the 2007-2008 world food price crisis, the Coalition for African Rice Development (CARD) was started in 2008 as an international agency initiative to promote rice agriculture in 23 SSA countries. Africa has an estimated 20 million ha of unexploited agricultural land suitable for rice cropping[14], along with vast untapped water resources for agricultural use[15]. The first-phase goal of CARD was to double rice production between 2008 and 2018, and this was achieved according to the Food and Agricultural Organization database (Fig. 1). The target for the second phase, which started in 2019, is to double rice production once again and achieve rice self-sufficiency as a major staple food by 2030, involving nine new participating countries.

Increases in rice production in SSA have been attained largely by area expansion[16]. Harvested rice area in SSA increased by 68% from 2008 to 2018, reaching 16 million ha. This can be compared to a total rice cultivation area of 30 million ha in China, the world's largest rice producer. The expansion has been mainly in West Africa countries (Fig. 1). CARD member countries have also extensively improved farming efficiency through fertilizer input, irrigation, and cultivation types[17]. World Bank-funded irrigation and water management systems have allowed large areas of irrigated rice farming. Irrigation also often allows for double cropping (two harvests per year), increasing the inundation period. Organic amendments and high nitrogen fertilizers have also been applied to raise rice yields.

Current methane emission inventories underestimate African rice emission and its trend because they do not adequately account for the recent increases in both cultivated area and emission factors (methane emission per unit of cultivated area). For example, the widely used Emissions Database for Global Atmospheric Research (EDGAR; Ref[18]) uses a constant emission factor of 1.3 kg ha$^{-1}$ d$^{-1}$ for irrigated fields in Africa, following 2006 IPCC[19] Tier 1 guidelines. A recent worldwide field study of rice emissions accounting for different practices and environmental conditions (soil texture, planting methods, pre-season water status) finds a best-estimate emission factor of 2.54 kg ha$^{-1}$ d$^{-1}$ for irrigated fields in Africa without organic amendments[20], twice the 2006 IPCC value. EDGAR obtains country-level harvested rice area from FAOSTAT and splits it into fixed fractions of upland (42%), rainfed (21%), irrigated (17%), and deepwater (20%) cultivation using 1993-1994 data from the International Rice Research Institute (IRRI[21]). Upland rice is grown in aerated soils and thus not a source of methane, but the upland fraction of rice area in Africa has decreased in recent decades[22] and averages 30% in post-2008 Africa[17, 22, 24, 25]. In Nigeria, IRRI[21] allocated 51% of the rice area to upland, while the more recent Global Upland Area database for 2009 allocates only 29% (Ref[24]). The ensemble of global emission inventories compiled by the Global Carbon Project (GCP; Ref[3]) indicates rice emissions in Africa of 1.8 ± 0.7 Tg a$^{-1}$ in 2017, where the standard deviation reflects the variability between individual inventories. This accounts for only 2% of total continental emissions of 84 ± 14 Tg a$^{-1}$. The GCP gives a trend for African rice emission of only 0.04 ± 0.02 Tg a$^{-1}$ a$^{-1}$ from 2006 to 2017 (Ref[4]).

Here we improve the estimate of African rice emission and its trend to reflect these different factors. First, we use the baseline emission factor for irrigated fields of 2.54 ± 0.50 kg ha$^{-1}$ d$^{-1}$ from Nikolaisen et al[20] and increase it by 10% following 2019 IPCC[23] for countries reporting organic amendments to CARD. Second, we take the country-level harvested rice area from FAOSTAT and split it into upland (zero emission), rainfed lowland (0.54 × irrigated), irrigated, and deepwater (0.06 × irrigated) by applying country-specific information collected from the latest Global Upland Area database, the national rice development reports for 2008-2018 to CARD, and the Africa Rice Center[17, 22, 25-26], as compiled in Supplementary Table S1. Third, we use country-specific cultivation periods from the comprehensive global rice calendar RiceAtlas database, ranging from 90 days in Gambia to 181 days in Kenya, and assume fields to be dry (no emission) outside of these cultivation periods.

Fig. 2 shows our estimate of country-level rice emissions and their trends. Our estimate of total rice emissions in Africa for 2008-2018 is 2.6 ± 0.50 Tg a$^{-1}$, compared to 1.5 ± 0.60 Tg a$^{-1}$ for 2006-2017 in the GCP. We find a rice emission trend of 0.20 ± 0.04 Tg a$^{-1}$ a$^{-1}$ in Africa for 2008-2018, considerably higher than the GCP trend of 0.04 ± 0.02 Tg a$^{-1}$ a$^{-1}$ Our rice emission trend can further be compared to the top-down total emission trends of 0.65 ± 0.15 Tg a$^{-1}$ a$^{-1}$ for Africa and 2.8 ± 0.67 Tg a$^{-1}$ a$^{-1}$ for the globe reported by the GCP for 2006-2017. Our results thus imply that African rice emissions have contributed 31% of the recent trend in African emissions and 7% of the trend in global methane emissions.

The Global Methane Pledge aims to reduce methane emissions by 30% by 2030 from 2019 levels, but feeding the rapidly growing population in Africa is expected to drive more aggressive expansion and intensification of rice cultivation in the years ahead[17]. This may require even greater reduction of methane emissions from other sectors.

**Acknowledgement**: This work was funded by the NASA Carbon Monitoring System, and by the Harvard University Climate Change Solutions Fund (CCSF). We wish to thank Lena Höglund-Isaksson (IIASA, Austria) and Matthew Hayek (NYU, USA) for their advice on African rice.

**Data Availability**: The Food and Agriculture Organization database (FAOSTAT) is available at https://www.fao.org/faostat/en/#data/QCL; Methane emission estimates from the Global Carbon Project (GCP) are available at https://www.icos-cp.eu/GCP-CH4/2019.

**Author contributions**. ZC and DJJ contributed to the study conceptualization. ZC conducted the data and analysis with contributions from DJJ, NB, HL, and HN. ZC and DJJ wrote the paper with input from all authors.

**Competing interests**. The authors declare that they have no conflict of interest.

**Supplementary information**. Supplementary Table S1.

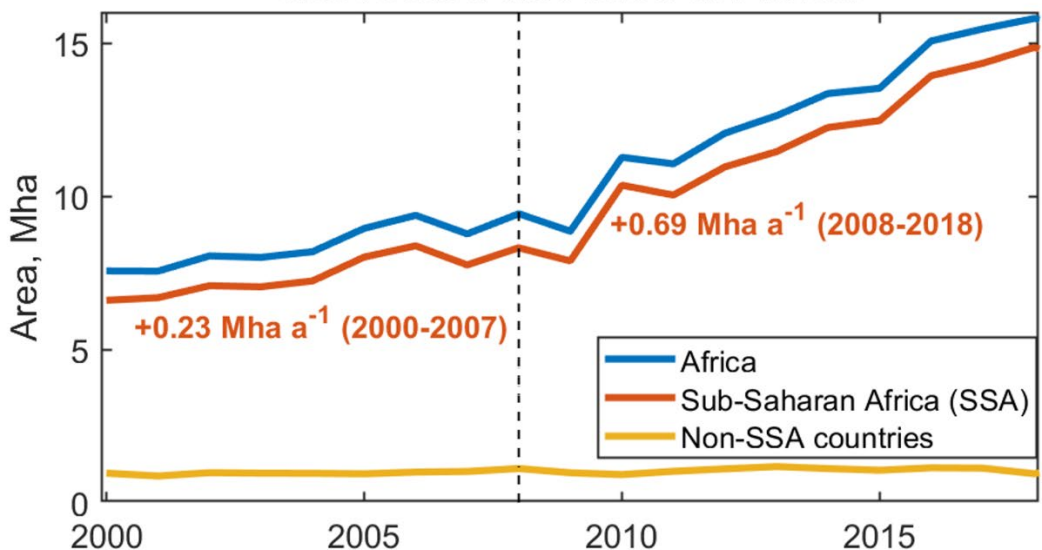
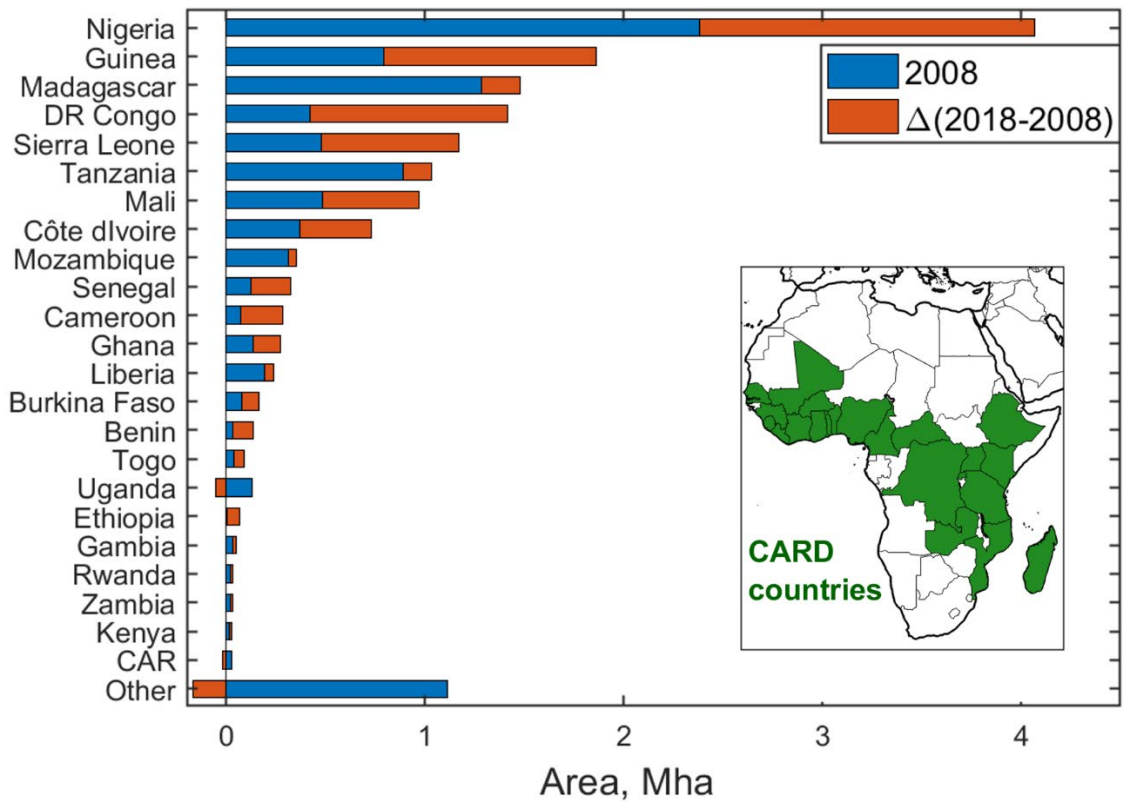

**Figure 1**. Harvested rice area in Africa based on FAOSTAT[27]. The top panel shows the rice area growth for 2000-2018. Black dotted line denotes 2008, the start year of the first phase of the Coalition for African Rice Development (CARD) program. Rice area growth rates in Sub-Saharan Africa (SSA) are given in legend, including the 23 CARD member countries as shaded in the inset map. The bottom panel shows rice area growth between 2008 and 2018 for each of the 23 CARD countries and for the rest of Africa (Other).

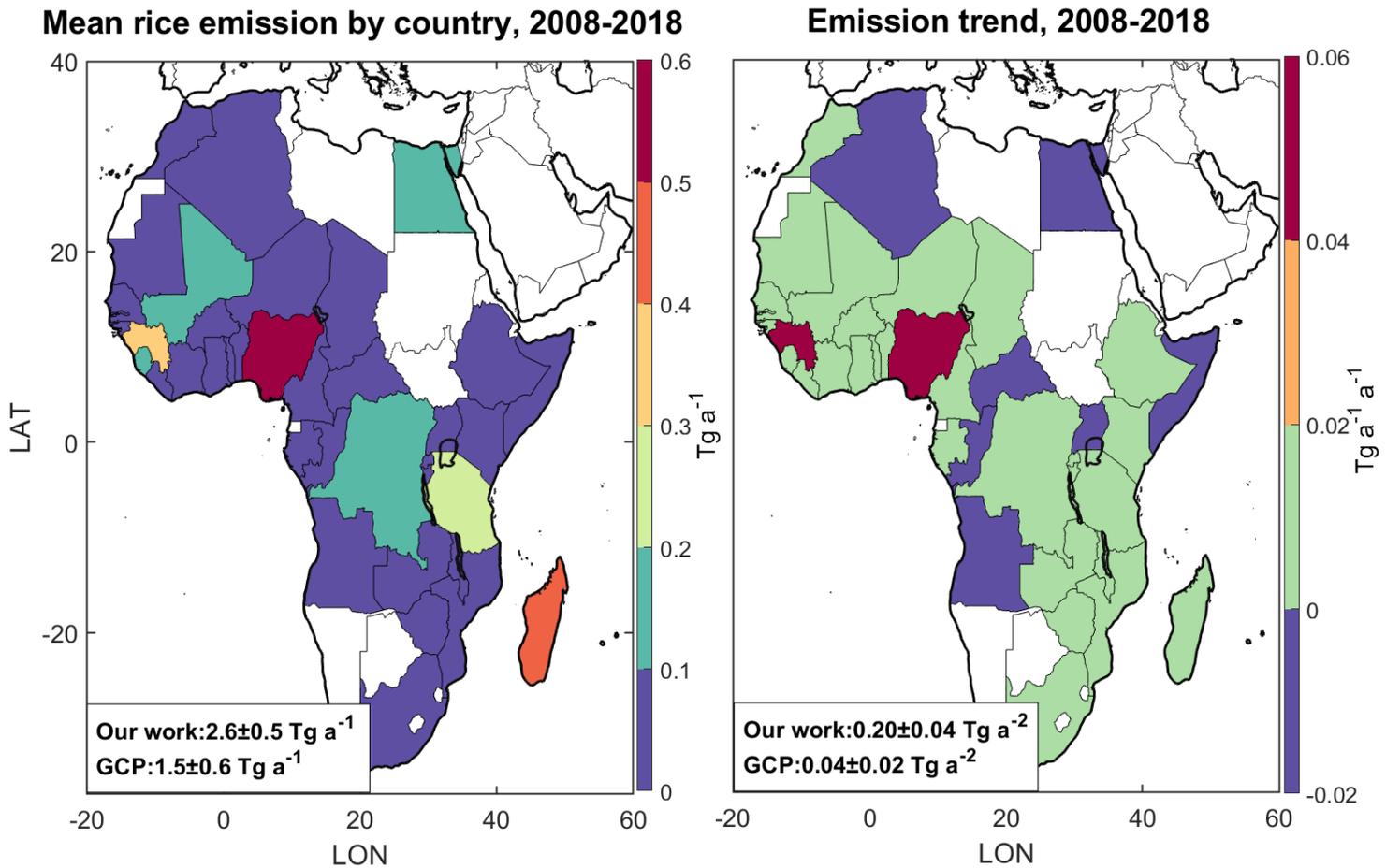

**Figure 2**. Country-level rice emissions and their trends in Africa for 2008-2018. The insets give the continental rice emissions and emission trends. GCP indicates the ensemble of inventories compiled in the Global Carbon Project for 2006-2017 (Ref[4]). Blank color indicates countries where rice area activity data are unavailable for 2008-2018 in FAOSTAT.